\pdfoutput=1
\newcommand{\green}{f_{_{\rm G}}}

\documentclass[fleqn,usenatbib,usedcolumn]{mnras}
\usepackage[british]{babel}             
\usepackage{newtxtext,newtxmath}

\usepackage[T1]{fontenc}
\usepackage{ae,aecompl}

\usepackage{graphicx}	
\usepackage{amsmath}	
\usepackage{amssymb}	
\usepackage{textcomp}
\usepackage{breakurl} 
\usepackage{textcomp}
\usepackage{subfig}
\usepackage{natbib}
\usepackage{epsf}
\usepackage{epstopdf}
\usepackage{caption}
\usepackage{color}
\usepackage{geometry}
\usepackage{fancyhdr}
\usepackage{transparent}
\usepackage[flushleft]{threeparttable} 
\usepackage{booktabs,caption}
\usepackage{mathrsfs}
\usepackage[margin=1cm]{caption}
\hypersetup{pdfauthor={T.James , P.Subramanian},
               pdftitle={Energetics of small electron acceleration episodes in the solar corona from radio noise storm observations},
               pdfkeywords={Solar-radio ,Solar-coronal heating,Solar-particle emission},
               bookmarksnumbered=true}
\let\la=\lesssim 

\title[Energetics of noise storms.]{Energetics of small electron acceleration episodes in the solar corona from radio noise storm observations}

\author[T.James et al.]{
Tomin James,$^{1}$\thanks{E-mail: tomin.james@students.iiserpune.ac.in}
Prasad Subramanian,$^{1,2}$
\\
$^{1}$Indian Institute of Science Education and Research, Pune\\
$^{2}$Center for excellence in Space Sciences (CESSI),Kolkata}

\date{Accepted XXX. Received YYY; in original form ZZZ}

\pubyear{2018}
\volume{{\rm in press}}

\begin{document}
\label{firstpage}
\pagerange{\pageref{firstpage}--\pageref{lastpage}}
\maketitle

\begin{abstract}
Observations of radio noise storms can act as sensitive probes of nonthermal electrons produced in small acceleration events in the 
solar corona. We use data from noise storm episodes observed jointly by
the Giant Metrewave Radio Telescope (GMRT) and the Nancay Radioheliograph (NRH) to study characteristics of the nonthermal electrons involved in the emission.
We find that the electrons carry $10^{21}$ to $10^{24}$ erg/s, and that the energy contained in the electrons producing a representative noise storm burst ranges from $10^{20}$ to $10^{23}$ ergs. 
These results are a direct probe of the energetics involved in ubiquitous, small-scale electron acceleration episodes in the corona, and could be relevant to a 
nanoflare-like scenario for coronal heating.
\end{abstract}




\section{Introduction}
\subsection{Small electron acceleration/heating events in the corona}
Decades after it was discovered that the solar corona is at a temperature in excess of a million K,
there is no consensus on why this is so, despite the fact that it is situated well
above the substantially cooler ($\approx 6000$ K) photosphere. One of the
ideas that has held considerable appeal is the concept of small events of electron acceleration and heating arising out of the continuous shuffling
and braiding of coronal magnetic fields. \citet{Parker_1988} outlined this idea and reasoned that the magnetic energy liberated in each episode is $\approx 10^{24}$ ergs,
which is around 9 orders of magnitude smaller than that liberated in a typical flare. Such small energy releases are very difficult to observe directly in thermal emission . On the other hand, bursts of radio emission at metric to decametric wavelengths are signatures of small populations of accelerated,
nonthermal electrons in the solar corona \citep{Ramesh_2012,Suresh_2017}. Even though these bursts involve miniscule numbers of nonthermal electrons,
the emission process is coherent, resulting in very high brightness temperatures, which makes them easily observable. Much of the research on nanoflares has concentrated on 
examining how an ensemble of such events can lead to observed temperature profiles in coronal loops 
\citep{Klimchuk_2001,Ishikawa_2017,Tripathi_2010,Testa_2014}. There have also been several 
MHD and hybrid particle/MHD simulation studies demonstrating how continually tangling coronal magnetic fields can provide potential sites for magnetic reconnection where
electrons can potentially be accelerated and account for local plasma heating
\citep{Arzner_2004,Sharma_2017,Guerreiro_2015, Guerreiro_2017,knizhnik_2018}. Electrons accelerated 
in such small reconnection episodes can often escape from the corona along open field lines, emitting type III radio emission and be detected in-situ by near-earth spacecraft
as impulsive electron bursts \citep{James_2017}. 
\subsection{Radio noise storms - sites of electron acceleration}
In this paper, we focus on radio noise storms, which are the archetypal example of electrons accelerated in 
small episodes outside of large flares  \citep{Klein_1995,klein_1998}. This kind of emission is believed to be caused by 
accelerated electrons trapped in coronal arches above active regions. Studying the energetics of nonthermal electrons 
responsible for radio noise storms provides a unique way of probing episodes of small scale electron acceleration and 
their possible relevance to the problem of coronal heating.

\noindent
The first reported detection of solar radio noise storms was probably during the second world war in 1942 by English radar stations.
They were wrongly interpreted as enemy attempts to jam radar signals, and it was only subsequently recognized that they originated from the solar corona \citep{Hey_1946}.
Noise storms (also referred to as type I emission) consist of a long-lived broad-band continuum ($\Delta f / f \sim 1 $) together with narrowband bursts.
Type I bursts have durations $\leqslant$ 0.5 s and bandwidths of $\Delta f / f \sim 3 \% $. Although noise storms have been studied extensively using whole Sun
dynamic spectra \citep{Elgaroy2016,kerdraon1979,Thejappa1991} there are not many instances where they have been imaged. \citet{Mercier_2006,Mercier_2015} combined data from 
the Nancay Radioheliograph (NRH) with data from the Giant Metrewave Radio Telescope (GMRT) to obtain meter wavelength images of unprecedented resolution and fidelity. In this paper,
we will use information on the fine structure of noise storm events from NRH-GMRT images reported by \citet{Mercier_2015}, together with high time resolution lightcurves 
from the NRH. We will use parameters derived from these detailed observations in the particle acceleration framework of \citep{Subramanian_2004,Subramanian_2006} to derive 
the fraction of nonthermal electrons (as compared to the thermal populations) that are involved in the noise storm emission process. We will also estimate the power input 
to the accelerated electrons that produce the noise storm emission and compare it with the heating requirements in the corona.

\section{Data Analysis}
\label{sec:dataanal}
\subsection{Noise storm lightcurves}
\begin{figure*}
	
	\begin{tabular}{cc}
	        \hspace{-2.0cm}
		\subfloat[The light curve for the noise storm event of Aug 27 2002 with the local maxima and minima marked.
		          We use a differential peak finder algorithm to locate the maxima and minima.]
		         {\includegraphics[scale=0.675]{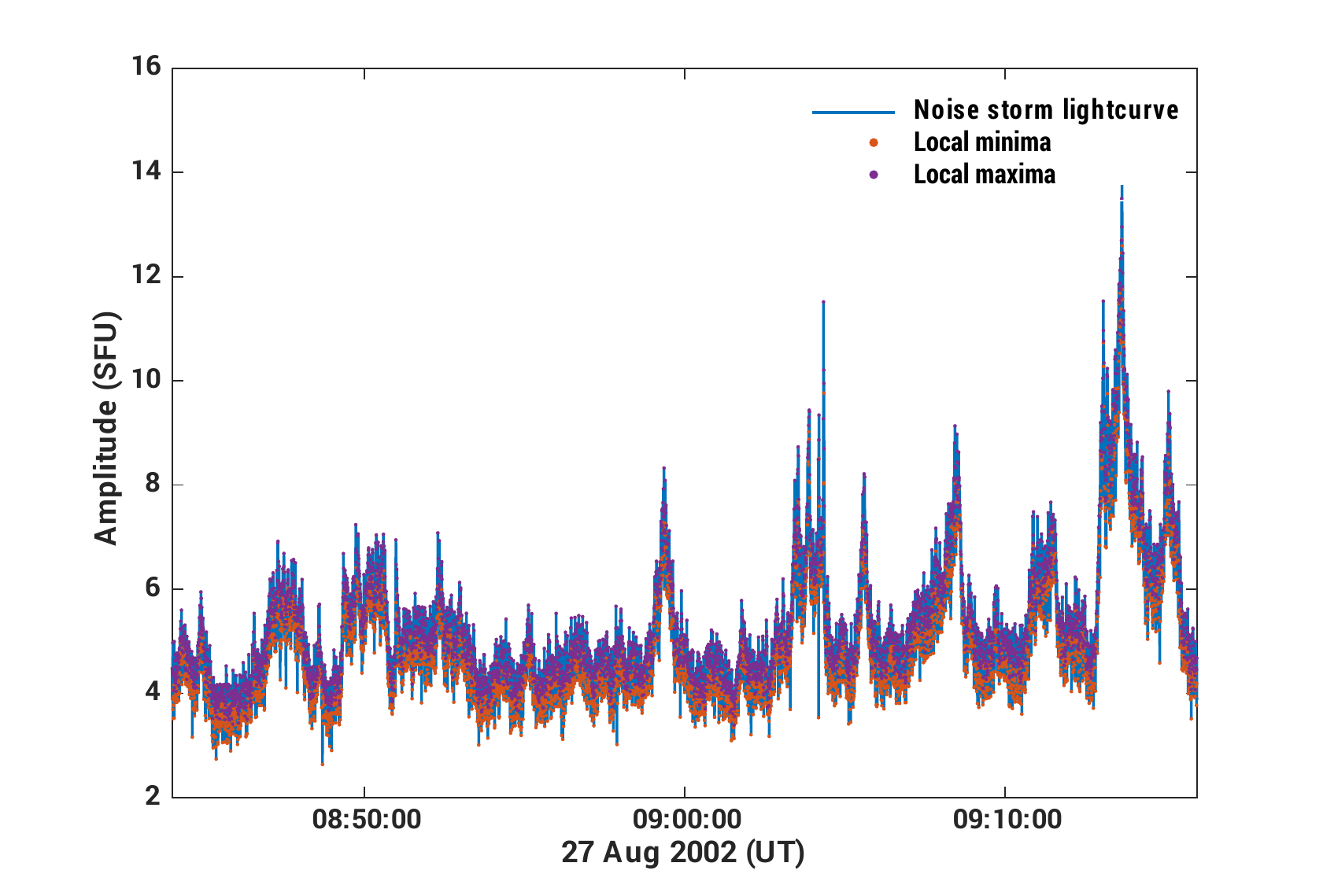}} &
		\hspace{-1.8cm}
		\subfloat[The locations of local minima in Fig 1a are used to construct the background trendline, which is marked in black.]
		{\includegraphics[scale=0.64]{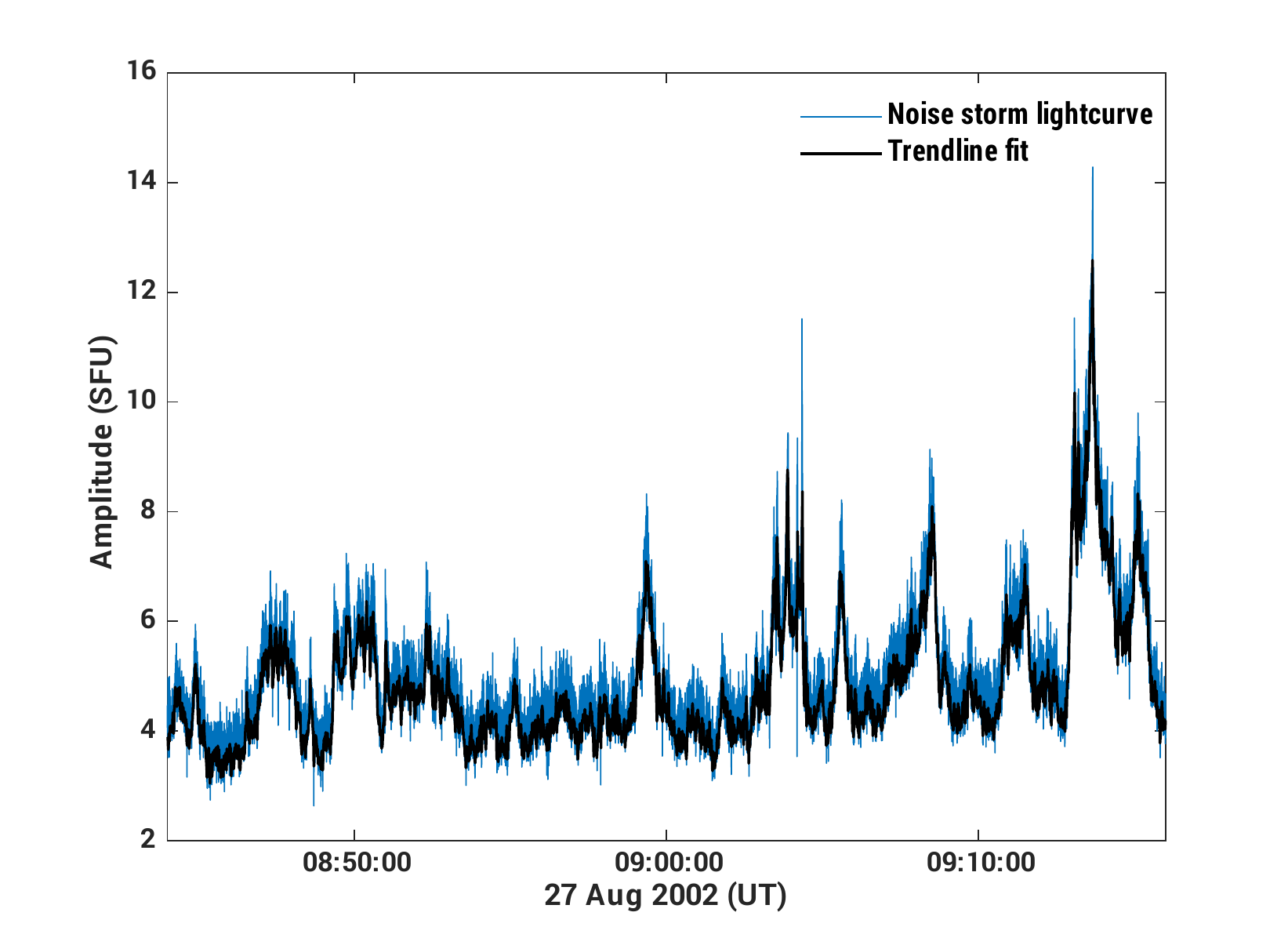}} \\
		\hspace{-2.0cm}
		\subfloat[A closeup of the lightcurve displayed in Fig 1b. It is evident that the trendline accurately traces both global and local features.]
		{\includegraphics[scale=0.64]{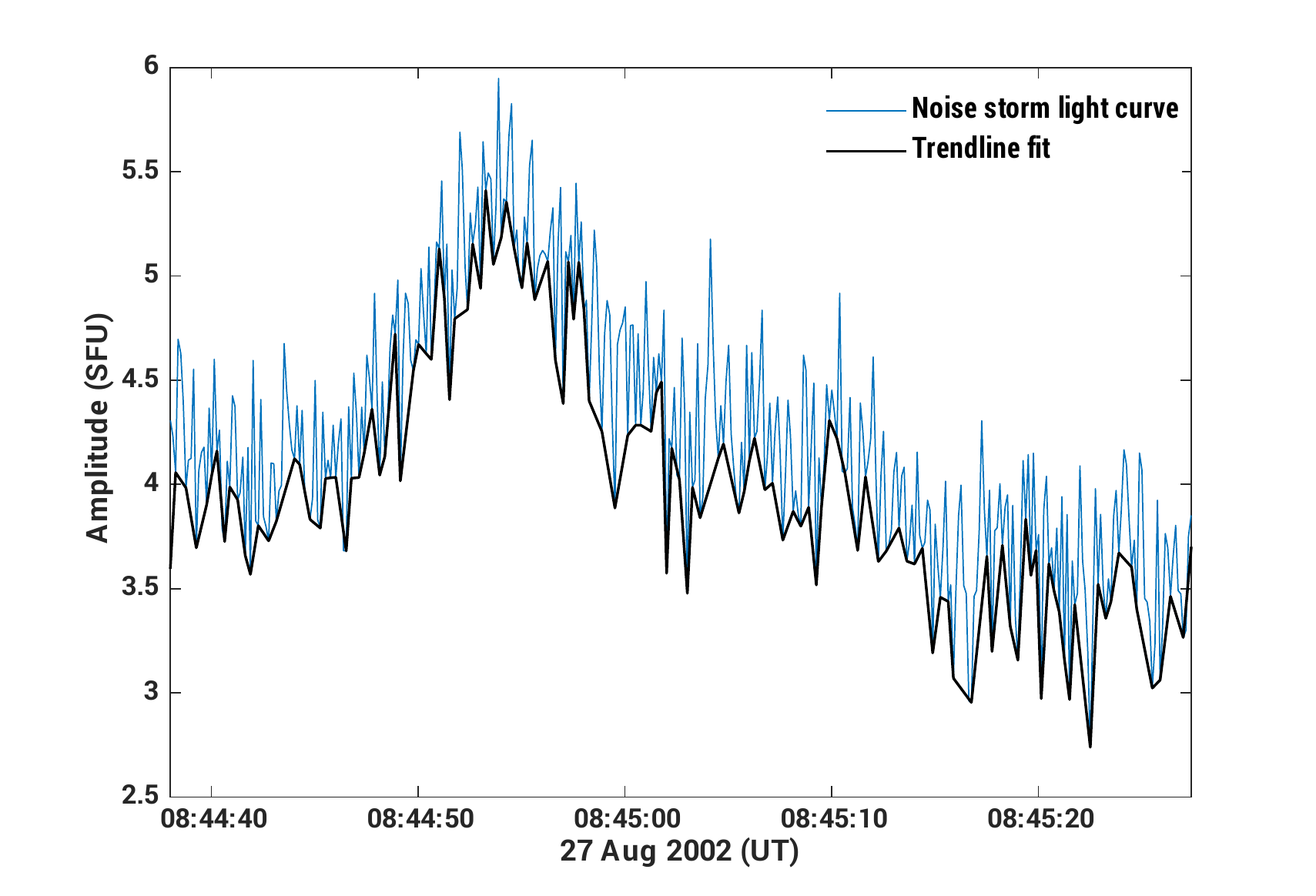}} &
		\hspace{-1.7cm}
		\subfloat[The background-subtracted lightcurve with gaussian fits to each of the burst overplotted]{\includegraphics[scale=0.648]{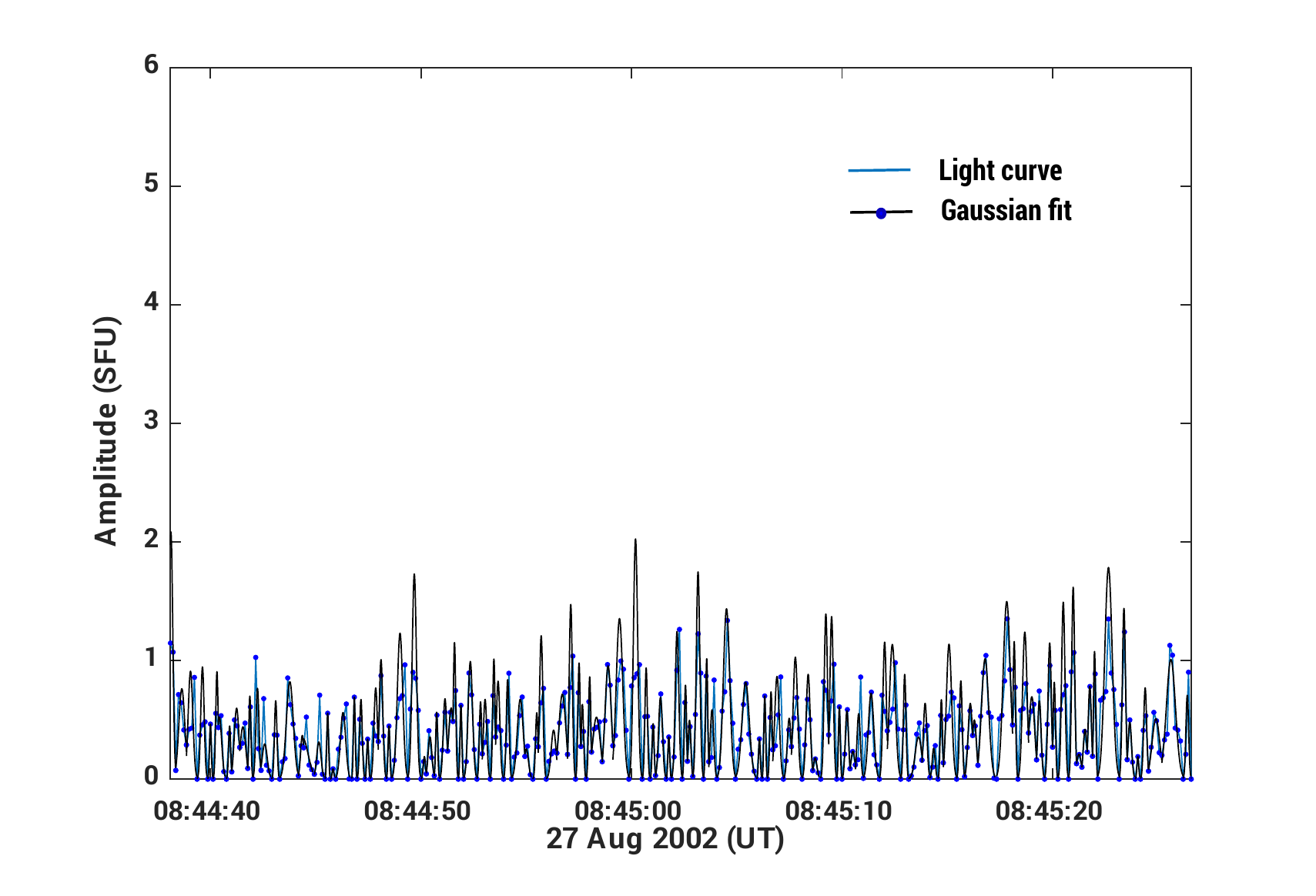}} \\

	\end{tabular}
	\caption{Background subtracted spikes for all the events are shown. Each burst is fitted with a Gaussian to  extract the peak amplitude, burst width and inter-burst times.}
	\label{fig:back_sub_proc}
\end{figure*}

In this paper, we concentrate on the four well observed noise storm events reported by \citet{Mercier_2015} using data from the NRH and the GMRT.
They include 327 MHz observations from 08:44 - 09:15 UT on Aug 27 2002, 327 MHz observations from  11:39 - 12:15 UT on Aug 14 2004, 236 MHz observations from 10:44 - 11:15 UT
on Jul 15 2003 and from 11:34 - 12:03 UT Apr 06 2006. The high resolution, high dynamic range images from these joint observations reveal the details of the fine spatial structure of the emission region(s),
together with any extended emission they might be embedded in. We refer the reader to \citet{Mercier_2015} for further details. The NRH-GMRT high resolution images give us
reliable estimates of the smallest source size in each instance. However, the integration times for these images were limited by the GMRT, which can be as long as 2 - 17 seconds.
We therefore use high time resolution data (time cadence of 0.125 seconds) from the NRH alone to study the detailed temporal behavior of the noise storm sources. In each instance,
we concentrate on the compact noise storm source reported in \citet{Mercier_2015} and use (high time resolution NRH) data corresponding only to the observation frequency and time 
range for which the joint NRH-GMRT observations were carried out. The details are listed in table \ref{tab:event_list}. The 327 MHz noise storm lightcurve from our observations on Aug 27 2002 is shown in Fig \ref{fig:back_sub_proc} as an illustrative example. As is typical for
noise storms, the high time resolution lightcurves reveal a series of spiky, short duration bursts superposed on a (relatively) slowly varying continuum envelope (Fig \ref{fig:back_sub_proc}). In each
instance, we first subtract the slowly varying envelope to obtain the detrended lightcurves
shown in Figs \ref{fig:back_sub_proc}[d]. We start with a peak finding algorithm which marks the local maximums(hills) and minimums(valleys) (Fig \ref{fig:back_sub_proc}a).
We use the valley positions to fit a polynomial interpolation function to obtain the background trendline. By using this approach we were able to trace both the global and local
undulations of the light curve (Fig \ref{fig:back_sub_proc}(b,c)). Each point of the the burst is subtracted from this trendline to obtain the background subtracted lightcurve. We fit a
gaussian to each of the background subtracted bursts to obtain distributions of amplitude,burst widths, and time between two adjacent bursts (Fig \ref{fig:back_sub_proc}d). We use the
peak flux from these lightcurves, together with the smallest source size estimates from \citet{Mercier_2015} to obtain an upper limit on the
brightness temperature of the noise storm emitting source. This quantity is listed in Table \ref{tab:observed_para} for each noise storm event . The high values of the brightness temperatures are clear evidence of the nonthermal nature of the emission. 
\subsection{Type I bursts - histograms of burst amplitude, width and inter-burst time}
\begin{figure*}
	
	\begin{tabular}{cc}
		\hspace*{-1cm}
		\subfloat[2002 Aug.27]{\includegraphics[scale=0.7]{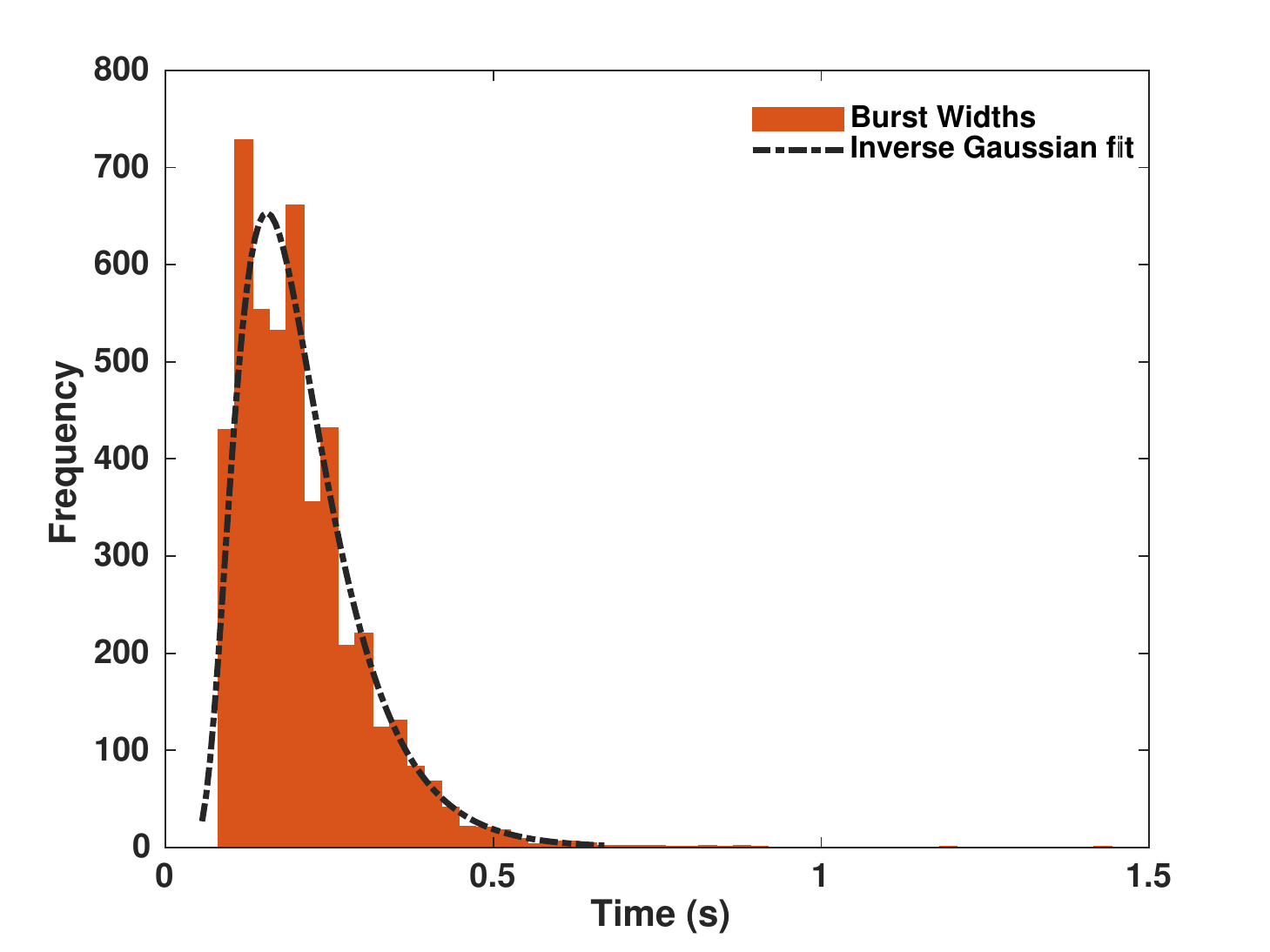}} &
		\hspace*{-1cm}
		\subfloat[2003 Jul.15]{\includegraphics[scale=0.7]{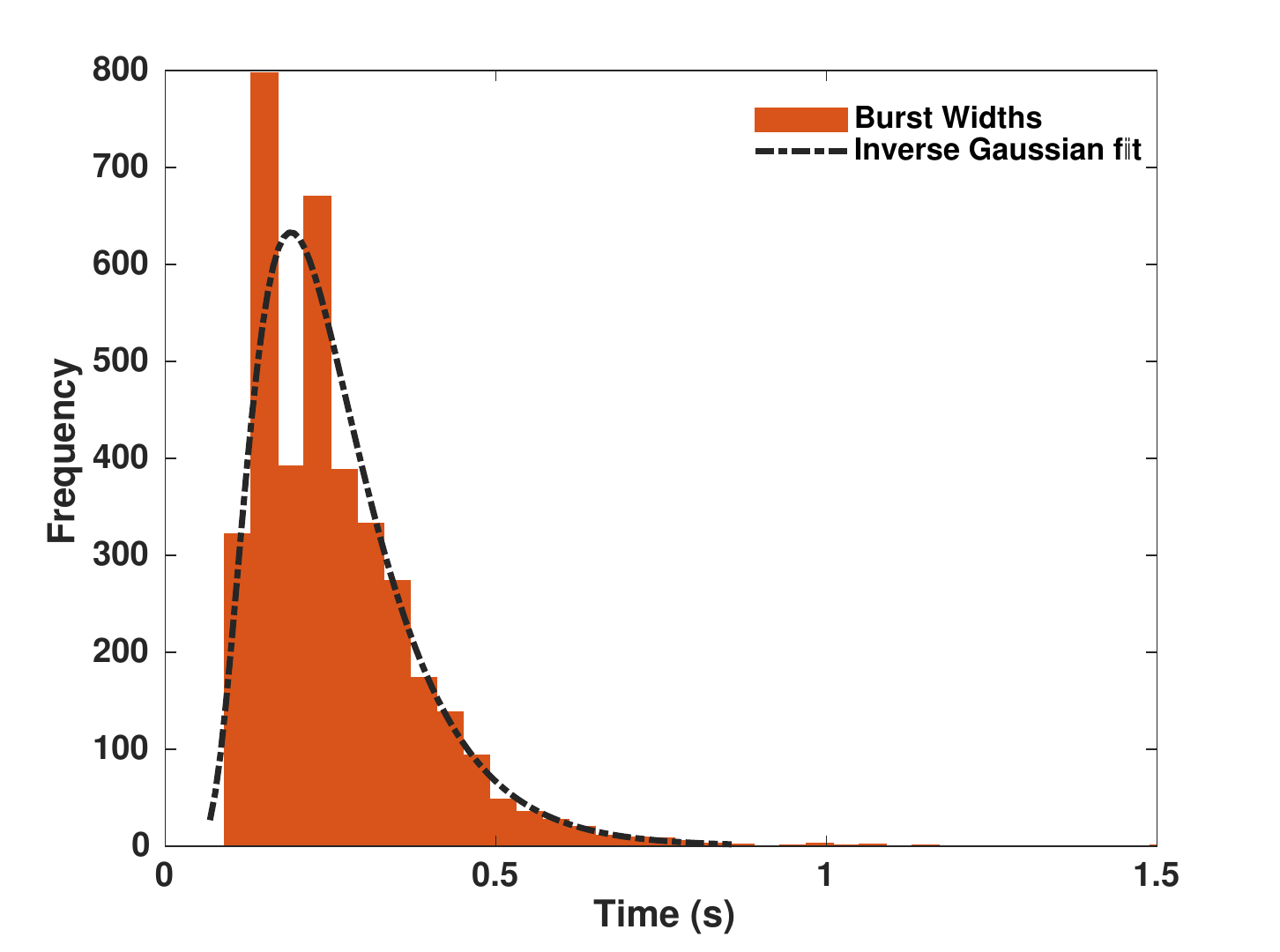}} \\
		\hspace*{-1cm}
		\subfloat[2004 Aug.14]{\includegraphics[scale=0.7]{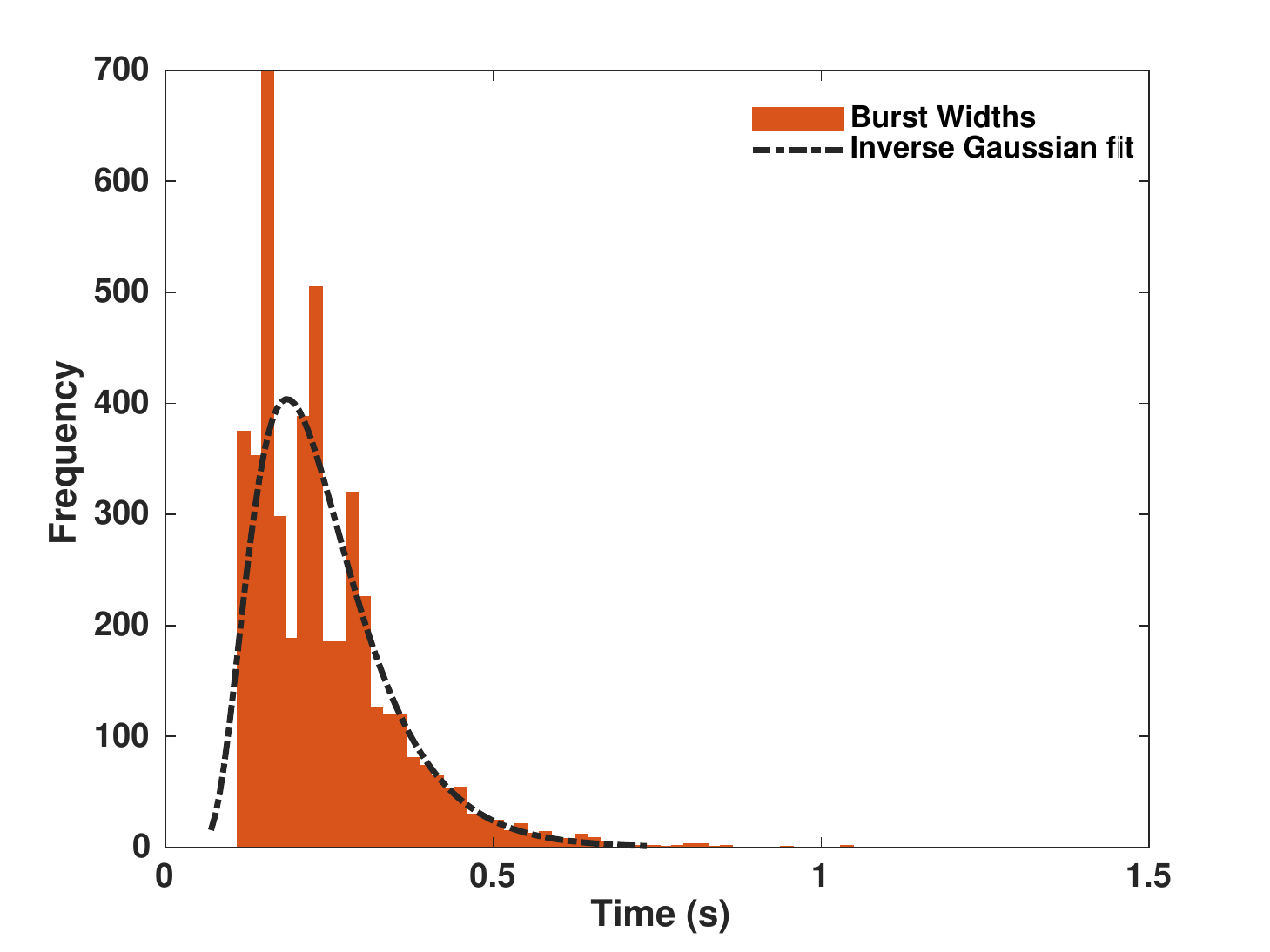}} &
		\hspace*{-1cm}
		\subfloat[2006 Apr.06]{\includegraphics[scale=0.7]{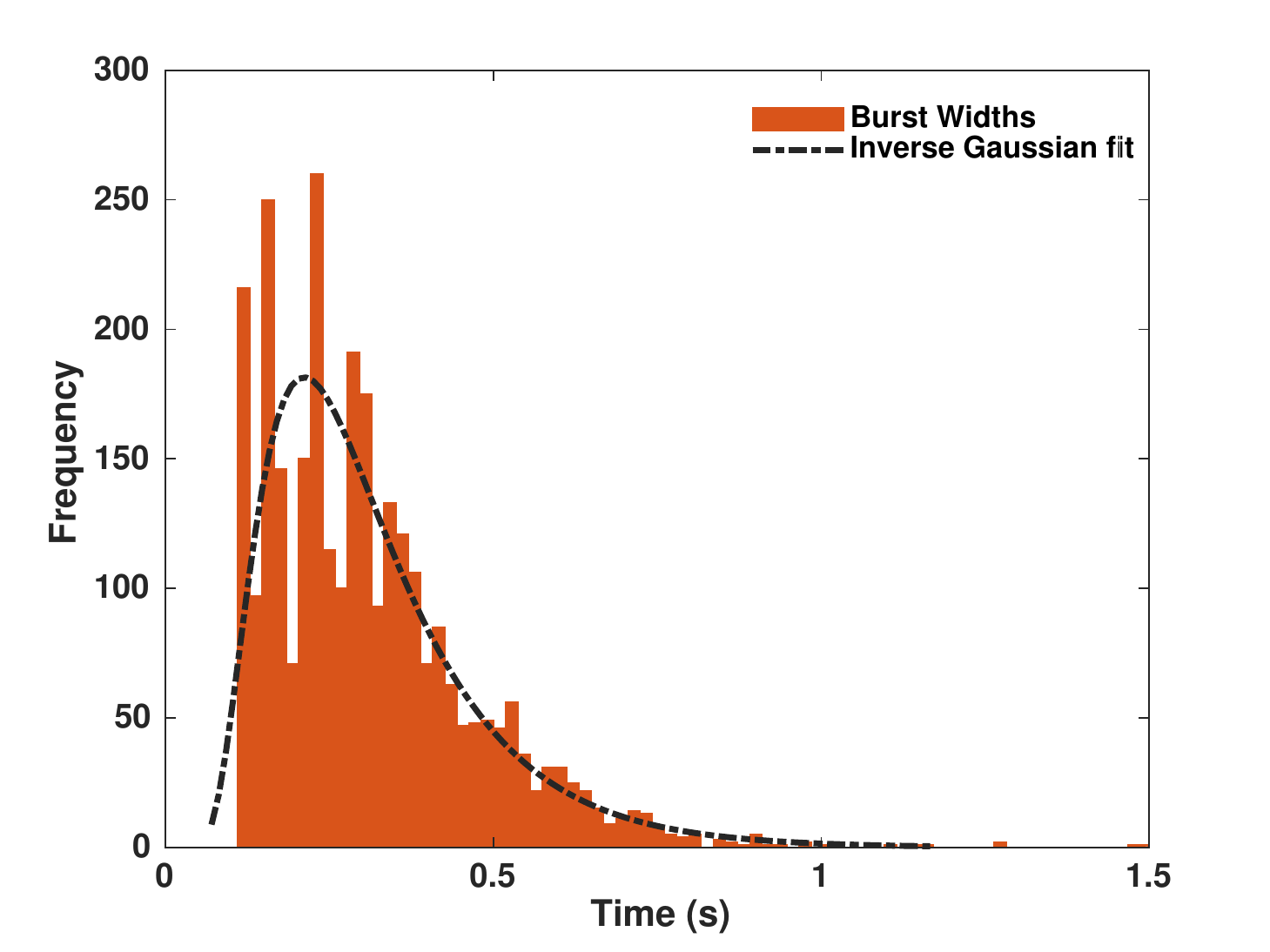}} \\

	\end{tabular}
	\caption{Histograms for burst widths. The widths are obtained by fitting a gaussian to each burst in the noise storm lightcurve.}
	\label{fig:hist_burst_width}
\end{figure*}

\begin{figure*}
	
	\begin{tabular}{cc}
		\hspace*{-1cm}
		\subfloat[2002 Aug.27]{\includegraphics[scale=0.7]{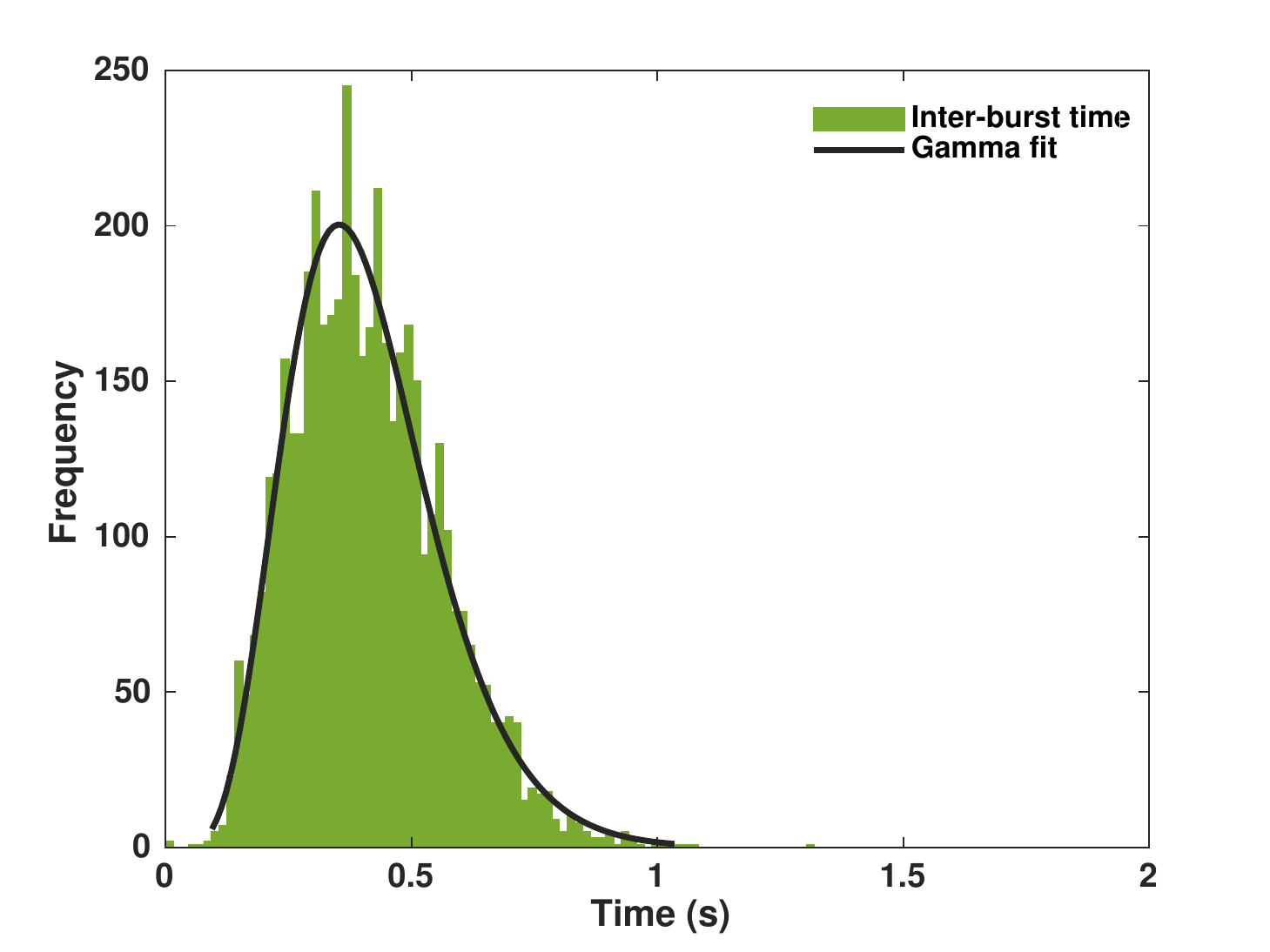}} &
		\hspace*{-1cm}
		\subfloat[2003 Jul.15]{\includegraphics[scale=0.7]{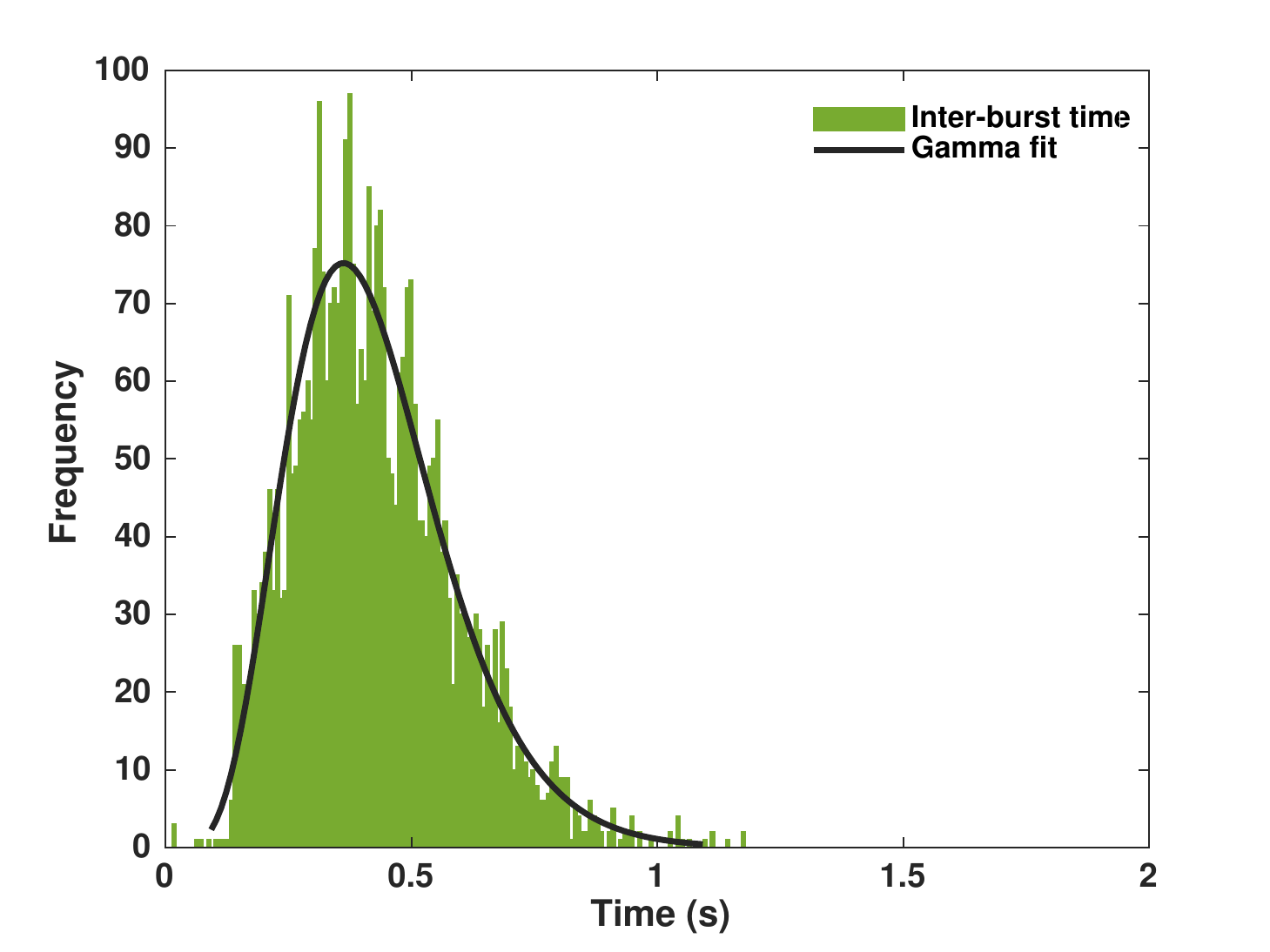}} \\
		\hspace*{-1cm}
		\subfloat[2004 Aug.14]{\includegraphics[scale=0.7]{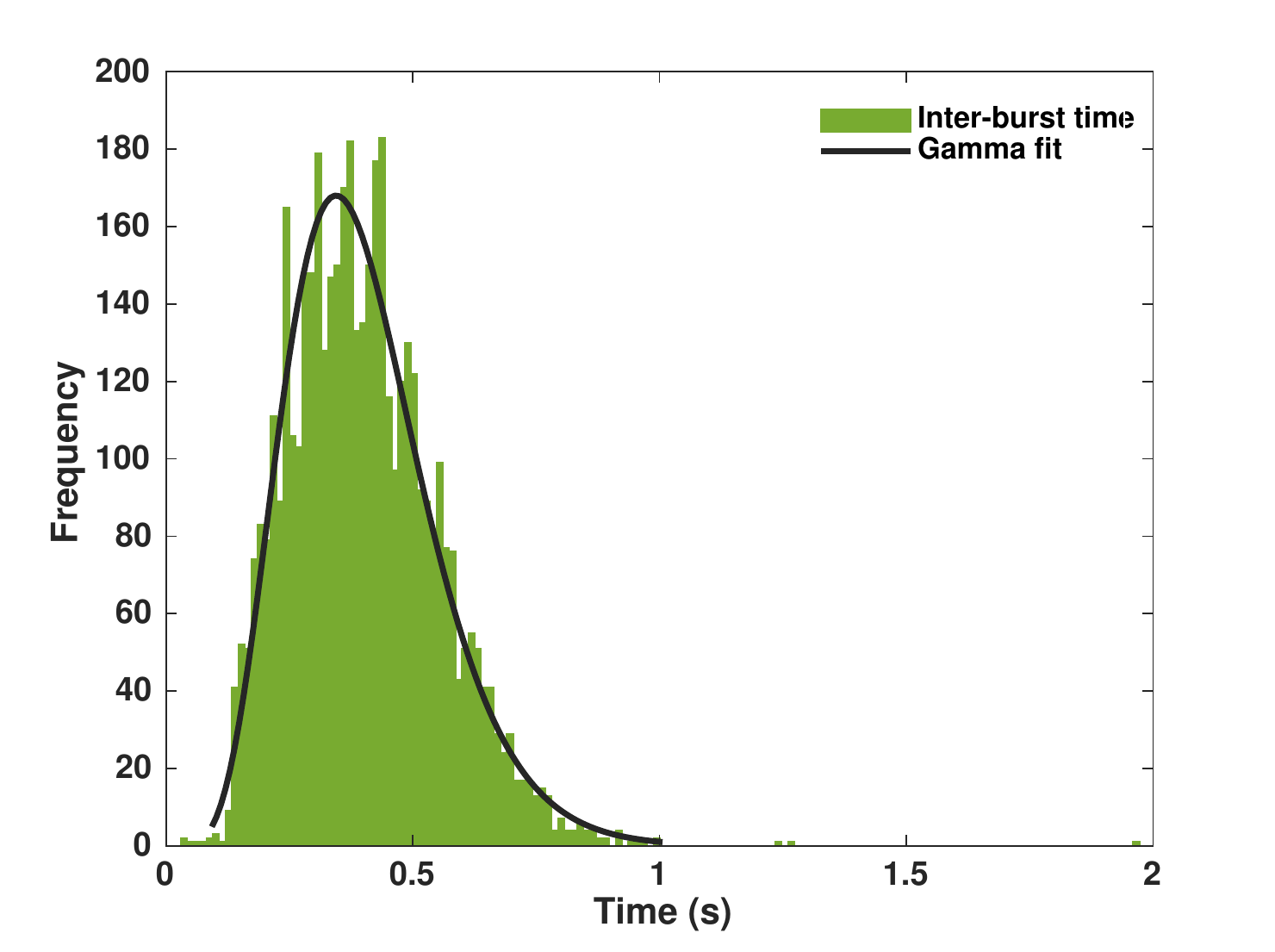}} &
		\hspace*{-1cm}
		\subfloat[2006 Apr.06]{\includegraphics[scale=0.7]{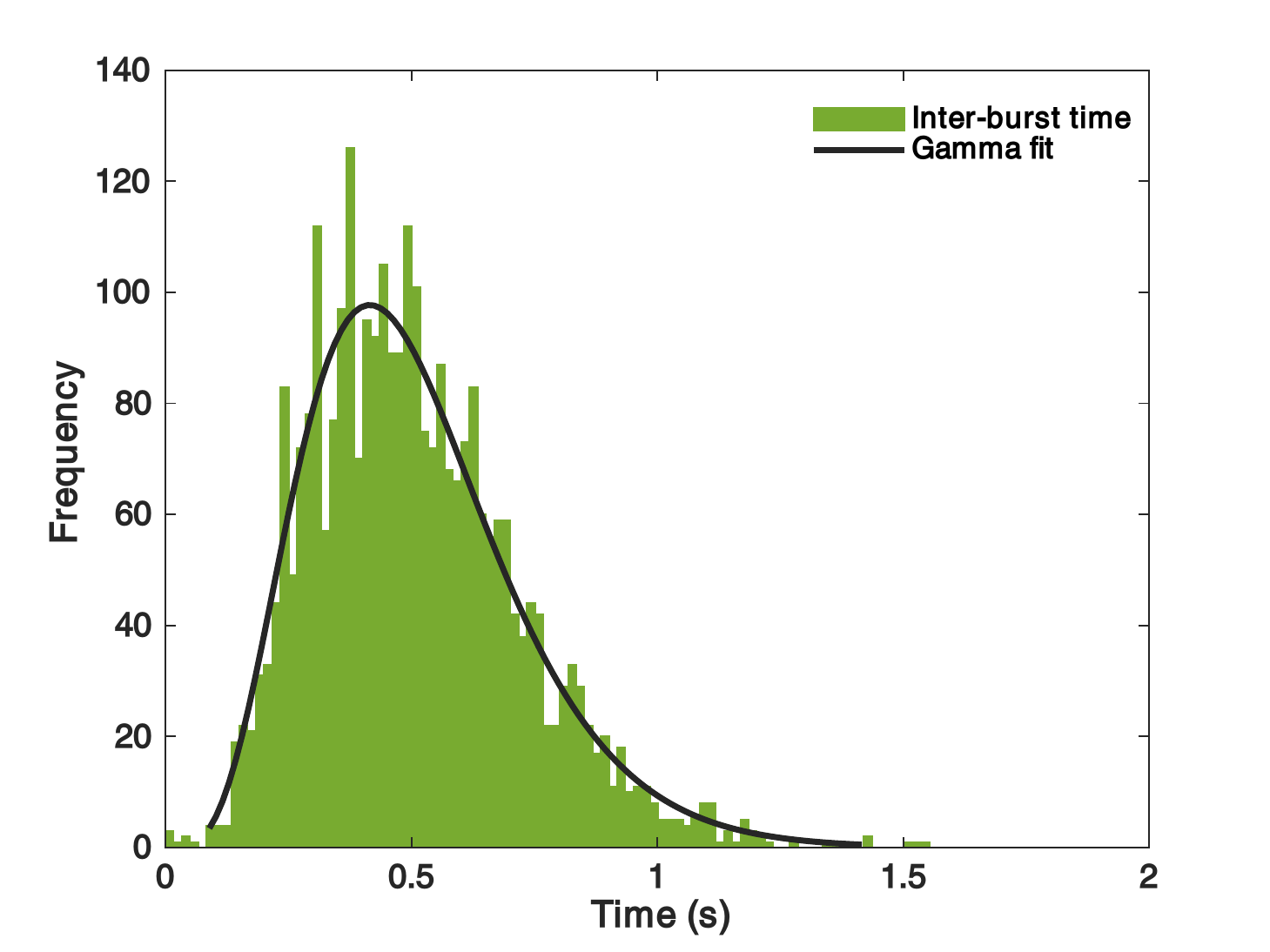}} \\
		
	\end{tabular}
	\caption{Histograms of the inter-burst duration. The Inter-burst duration is defined as the time duration between two adjacent peaks in the noise storm lightcurve.}
	\label{fig:hist_inter}
\end{figure*}

In order to identify type I bursts from the lightcurves, we identify peaks by finding zero crossings in the slope. An individual burst is defined as one which has a peak and two adjoining valleys.  We fit a Gaussian function to each such burst to determine 
its amplitude and width. This enables us to construct frequency histograms of burst amplitudes, widths and inter-burst separation for each observation (Fig \ref{fig:hist_burst_width} and Fig 
\ref{fig:hist_inter}) .

\begin{figure*}
	
	\begin{tabular}{cc}
		\hspace*{-0.5cm}
		\subfloat[2002 Aug.27]{\includegraphics[scale=0.6]{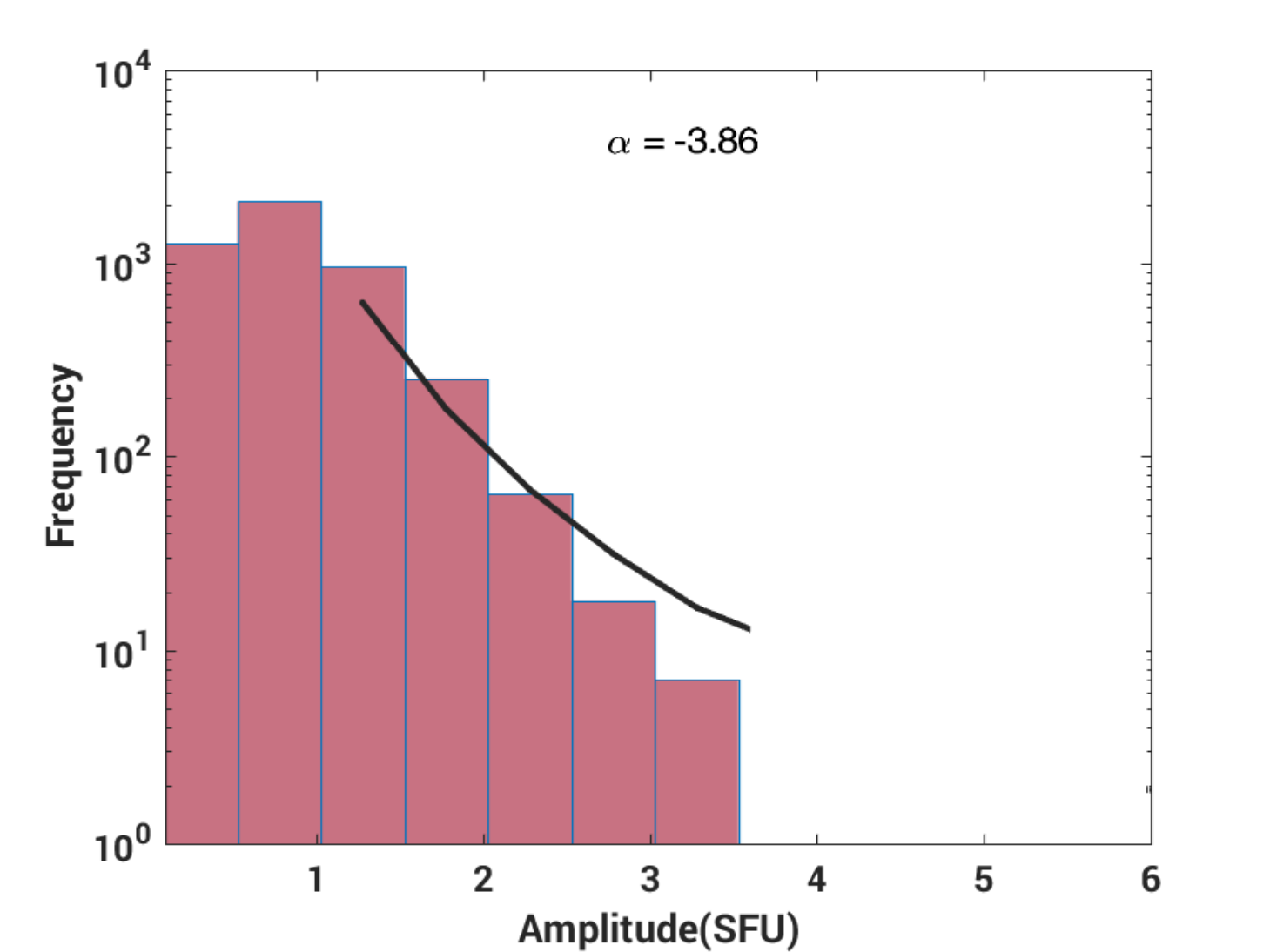}} &
		\hspace*{-0.5cm}
		\subfloat[2003 Jul.15]{\includegraphics[scale=0.6]{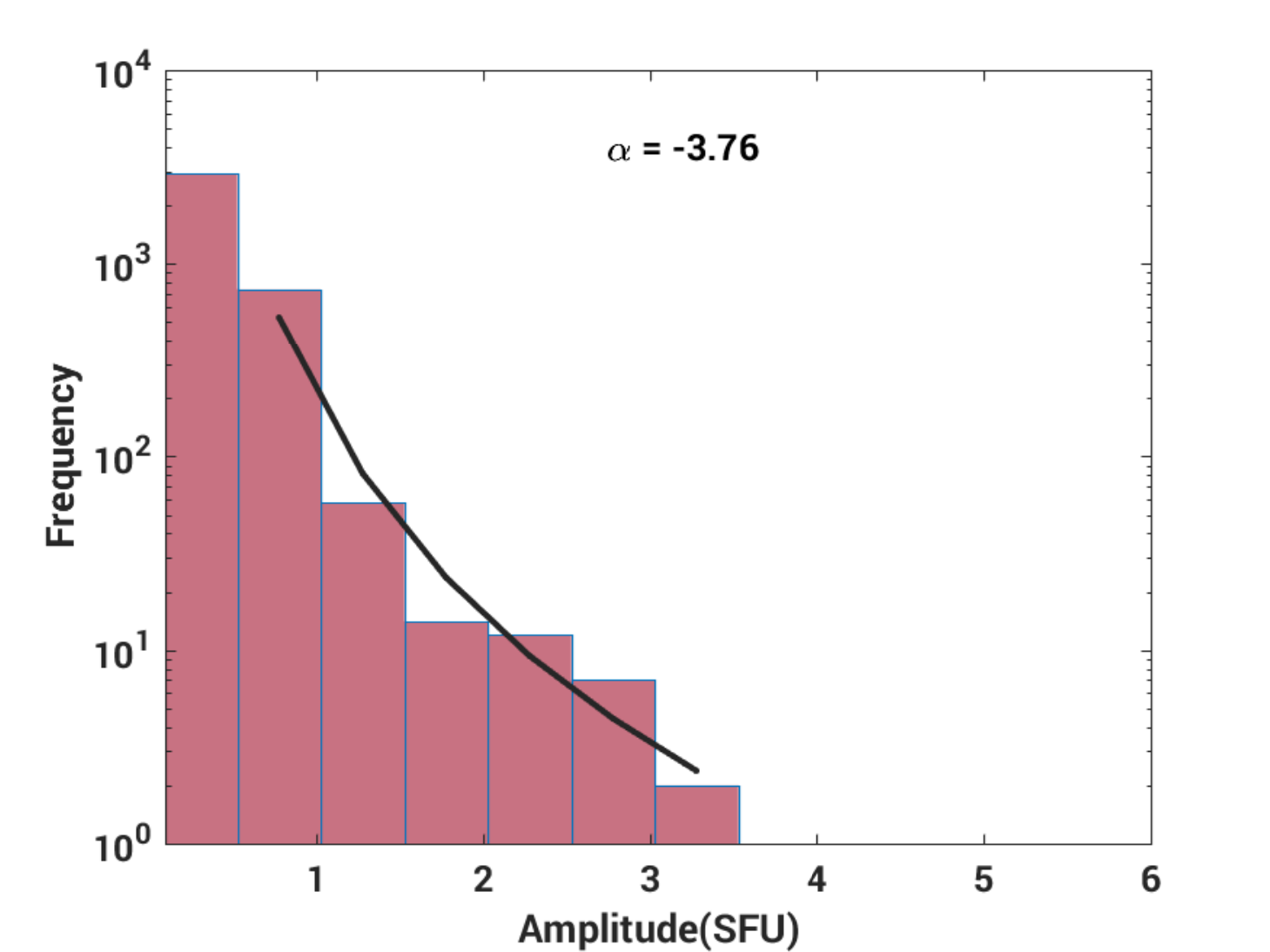}} \\
		\hspace*{-0.5cm}
		\subfloat[2004 Aug.14]{\includegraphics[scale=0.6]{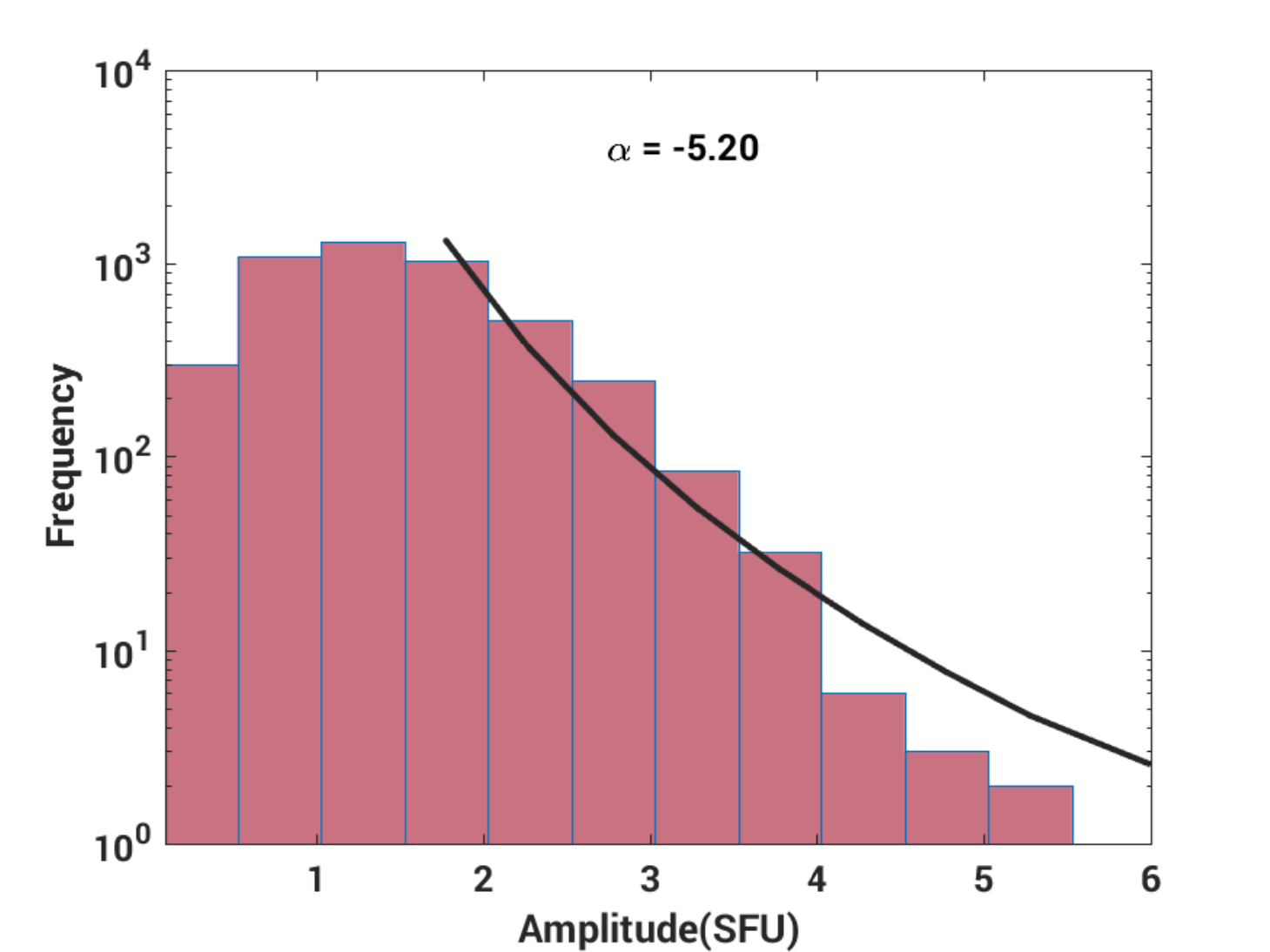}} &
		\hspace*{-0.5cm}
		\subfloat[2006 Apr.06]{\includegraphics[scale=0.6]{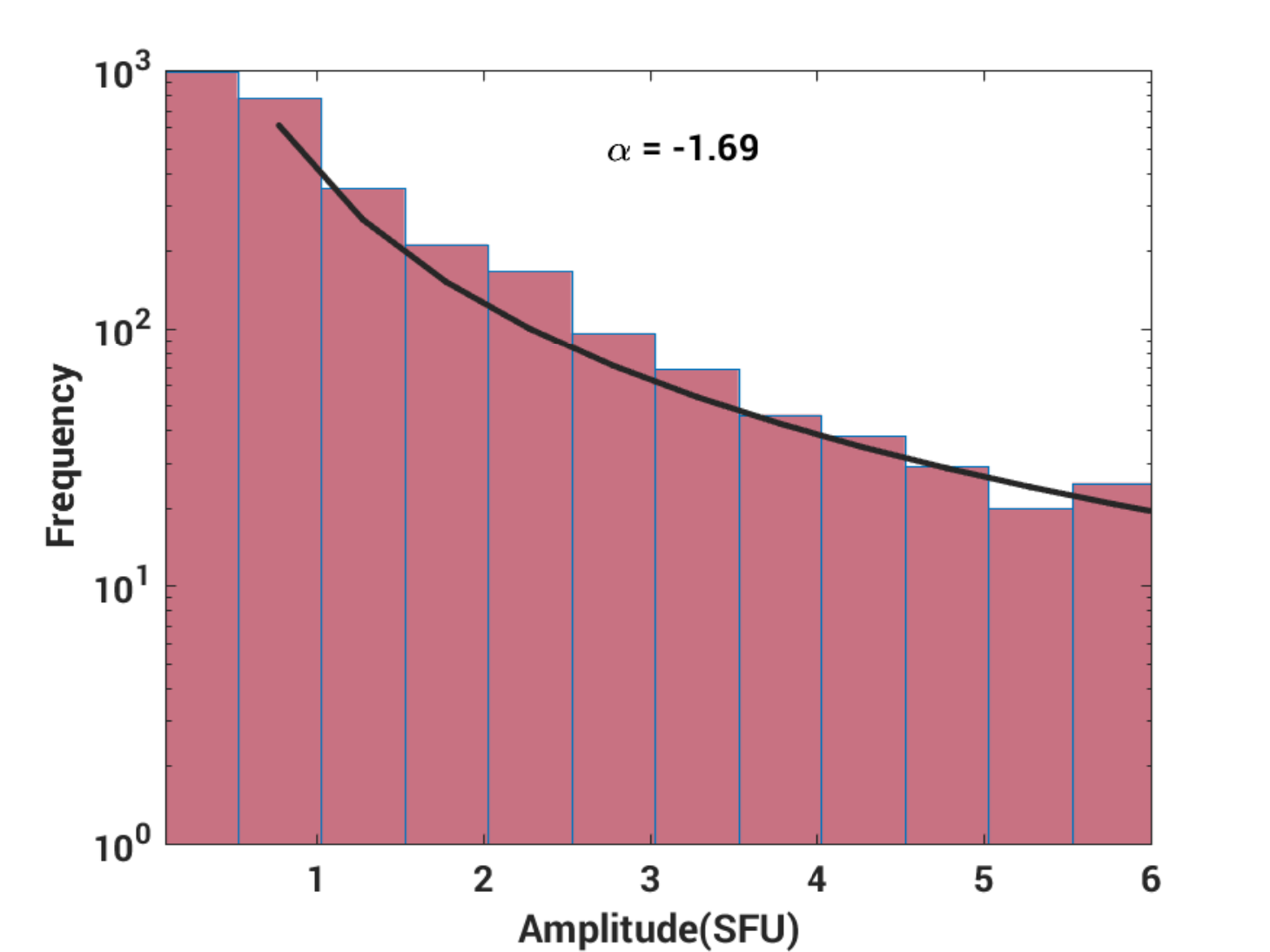}} \\
		
	\end{tabular}
	\caption{Histograms of occurrence frequency versus flux for each of the noise storm episodes we study. The power law fit is shown by the black curve.}
	\label{fig:powerlaw_amp}
\end{figure*}


The peak amplitude histograms we obtain are similar to the ones shown in \citet{Mercier_1997}. In each instance, we fit a a power law of the form $dN / dA \simeq A^{\alpha}$, where A 
is a measure of the amplitude, dN is the number of events recorded between A and A+dA and $\alpha$ is the slope of the $dN / dA$ curve on a log-log plot. The results are 
shown in Fig \ref{fig:powerlaw_amp}. The power law indices lie in the range $-5.2 < \alpha < -1.6$ (Table \ref{tab:observed_para}).
Strictly speaking, if these small scale heating events are the primary contributor to steady-state coronal heating, it is necessary that $\alpha < -2$ in order to avoid a situation where the 
total energy diverges \citep{Hudson_1991}. Our findings may be contrasted with the range $-3.5 < \alpha < -2.8$ found 
by \citet{Mercier_1997},\citet{Ramesh_2012} and \citet{Iwai2013,Iwai2014}. \citet{Suresh_2017} found that $\alpha = -2.23$ for very small energy releases observed with the Murchison Widefield Array. 
Recent simulations show that the energy released in small reconnection events in the corona also obey power
law statistics \citep{knizhnik_2018}.

Since we have fitted Gaussian profiles to each individual burst, we are in a position to construct histograms of burst widths (which is estimated from the Gaussian fits) and for 
the interval between bursts (which is defined as the interval between two successive peaks). These histograms are shown in Fig \ref{fig:hist_burst_width} and Fig 
\ref{fig:hist_inter}. In order to interpret the results
in terms of a theoretical framework for electron acceleration, we need average values for the burst duration and the inter-burst interval. The histograms for the burst width are somewhat skewed towards high values. An appropriate function to describe such a distribution is the inverse
Gaussian function:
\begin{equation}
f(x;\mu,\lambda)
= \left[\frac{\lambda}{2 \pi x^3}\right]^{1/2} \exp\left\{\frac{-\lambda (x-\mu)^2}{2 \mu^2 x}\right\}
\label{invgaus}
\end{equation}
Here, $\mu$ ($> 0$) is the mean and  $\lambda$ ($> 0$) is the shape parameter. The fits are denoted by the black lines 
superposed on the histograms in Fig \ref{fig:hist_burst_width}. On the other hand, the inter-burst time histograms are not as skewed, and an appropriate function to fit them is the Gamma distribution given by 
\begin{equation}
 f(x;\alpha,\beta) = \frac{ \beta^\alpha x^{\alpha-1} e^{-\beta x}}{\Gamma(\alpha)},
 \label{gamma_fit}
\end{equation}
where $\alpha$ ($> 0$) is the shape parameter, $\beta$ ($> 0$) is the scale parameter and $\Gamma(\alpha)$ is the complete Gamma function. The mean of this distribution is $\alpha/\beta$. The fits are shown in Fig 
\ref{fig:hist_inter}. The mean values of the burst-width and inter-burst duration are listed in columns 3 and 4 respectively of Table \ref{tab:observed_para}.

\section{Energetics of accelerated electrons}
\subsection{Background}
One of the main objectives of our work here is to use our observations to estimate 
the power in the accelerated electrons responsible for radio noise storm emission. 
This kind of emission is due to coherent plasma emission from accelerated (also frequently called nonthermal) electrons. 
The acceleration process responsible for producing these electrons is not very well understood; it could be due to en ensemble of weak
shocks produced by emerging flux \citep{Spicer_1982,Benz_1987} or due to acceleration in reconnection sites arising from magnetic 
flux emergence \citep{Bentley_2000, Li_2017}. In either scenario, the electrons are likely subject 
to a second order Fermi acceleration-like process. Reflections from shrinking magnetic islands produced during reconnection can also result in a first 
order Fermi-like acceleration process \citep{Drake_2006}. A typical noise storm emitting region contains a large number of acceleration regions. There are a large number of scattering centers within an acceleration region, and electrons gain energy over repeated interactions with these scattering centers. \citet{Subramanian_2004,Subramanian_2006} have applied a second order Fermi acceleration mechanism to analyze 
the energetics of electrons responsible for noise storm continua. However, they used the ratio of the acceleration timescale to the residence timescale as a free parameter. 
We use the high temporal resolution light curves to constrain the following important free parameters in the model: the acceleration timescale ($t_{\rm accel}$), which is interpreted as the time between bursts and the residence time of the electron ($\tau$) in the acceleration region,
which is interpreted as the burst duration.  The inter-burst timescale is representative of the lifetime of an individual acceleration region, while the duration of a burst is indicative of the time an electron spends in the acceleration region over the numerous collisions it suffers with scattering centers inside the acceleration region. The dimensions of the noise storm emission
region are determined from the high resolution NRH+GMRT images. 

For the sake of completeness, we recapitulate the essential features of the model from \citet{Subramanian_2004}.
The electrons are assumed to be subject to a generic stochastic acceleration process described by

\begin{equation}
{\partial \green \over \partial t} =
\frac{1}{p^2}\,\frac{\partial}{\partial p}\left( p^2 \, {\cal D}\,
\frac{\partial \green}{\partial p}\right) + {\dot N_0 \, \delta(p-p_0)
\over p_0^2} - {\green \over \tau} \ ,
\label{eq1}
\end{equation}

Equation~(\ref{eq1}) expresses the evolution of the Green\textquotesingle s function 
for the electron distribution ($\green$) as a consequence of diffusion in momentum space due to stochastic interactions with
scattering centres (first term on the right hand side), injection of electrons at a momentum $p_{0}$ (second term on the RHS) and escape from the
acceleration region (third term on the RHS).  Noise storms typically last for timespans ranging from a few hours to days. The emission comprises a series of short-lived bursts
(typical durations are a fraction of a second), superposed on a relatively steady continuum. We focus here on the steady-state solution of equation~(\ref{eq1}). In other words,
we concentrate on timescales much longer than acceleration or residence timescales; this can be taken to mean that we address the energetics of the noise storm continuum. 
Assuming that the properties of the continuum arise from a superposition of several individual bursts, we derive average properties of the individual bursts.

With a momentum diffusion coefficient given by \citep{Ptuskin1998,Chandran2004,Miller1996,Luo2003}
\begin{equation}
{\cal D} = D_0 \, p^2 \, ,
\label{eq2}
\end{equation}
the steady-state ($\partial/\partial t \rightarrow 0$) solution to the Green\textquotesingle s function is \citep{Subramanian_1999,Subramanian_2004,Subramanian_2006}.
\begin{equation}
\green(p,p_0) = A_0 \, \begin{cases}
(p/p_0)^{\alpha_1} \ , & p \le p_0 \ , \cr
\phantom{space} \cr
(p/p_0)^{\alpha_2} \ , & p \ge p_0 \ , \cr 
\end{cases}
\label{eq3}
\end{equation}

The exponents $\alpha_{1}$ and $\alpha_{2}$ are related to $D_{0}$ and $\tau$ via
\begin{equation}
\alpha_{1} \equiv -\frac{3}{2}+\Big(\frac{9}{4} + \frac{1}{D_{0}\tau}\Big)^{1/2}, \quad \alpha_{2} \equiv -\frac{3}{2}-\Big(\frac{9}{4} + \frac{1}{D_{0}\tau}\Big)^{1/2},
\label{eq4}
\end{equation}

The acceleration timescale is related to $D_{0}$ via
\begin{equation}
t_{\rm accel} = \frac{1}{8 D_{0}} \, .
\label{eq5}
\end{equation}

Equation~(\ref{eq3}) is a solution for the Green\textquotesingle s function (i.e., a solution for injection of electrons at a single momentum $p = p_{0}$ 
into the acceleration mechanism). It does not include losses, but we can expect net acceleration only when the acceleration timescale (Eq~\ref{eq5}) is 
shorter than the loss timescale. For electrons emitting Langmuir waves, the relevant loss timescale is the Coulomb loss timescale. Accordingly, net acceleration 
is possible only for $p >$ a critical momentum $p_{c} = 1.64 \times 10^{-21} \left(n_e / D_0\right)^{1/3}$, where $n_{e}$ is the background (thermal) electron density. Since we assume the observed emission to be (fundamental) plasma emission, the background electron density corresponds to the plasma level from which the emission originates.
 In other words, $n_{e} = (1/81) \times 10^{6} f^{2}\, {\rm cm^{-3}}$, where $f$ is the observing frequency in MHz. For an observing frequency of 327 MHz, $n_{e} = 1.3 \times 10^{9}\,{\rm cm}^{-3}$, while $n_{e} = 6.8 \times 10^{8}\,{\rm cm}^{-3}$ for an observing frequency of 236 MHz. Given that noise storms typically occur over active regions, these values are typical. Assuming that electrons from the tail of the thermal
 (Maxwellian) electron distribution (with $p > p_{c}$) are subjected to net acceleration, \citet{Subramanian_2004}
obtain the following expression for nonthermal electron distribution function ($f$), which is a convolution of the Green's function ($\green$) with the tail of the thermal
Maxwellian distribution for $p > p_{c}$:
\begin{equation}
f(p) =
{n_e \left\{
\xi^{\alpha_1/2} \, \Gamma\left(- \, {\alpha_1 \over 2} , \xi\right)
- \xi^{\alpha_2/2} \left[ \Gamma \left(- \, {\alpha_2 \over 2} , \xi
\right) - \Gamma \left(- \, {\alpha_2 \over 2} , \xi_c \right)
\right]\right\}
\over \sqrt{2 \pi} \, (m_e k T)^{3/2} \, (\alpha_1 - \alpha_2)
\, D_0 \, \tau}
\ ,
\label{particular}
\end{equation}
The moments of the distribution function ($f$) describing the nonthermal/accelerated electrons are as follow. The number density ($n_{*}$) of nonthermal electrons is given by
\begin{equation}
n_{*} \, ({\rm cm^{-3}}) \, = \int_{p_{c}}^{\infty} p^{2} f(p) dp
\label{numdens}
\end{equation}
and its energy density is
\begin{equation}
U_{*} \, ({\rm erg\,cm^{-3}}) \, = \int_{p_{c}}^{\infty} \epsilon \, p^{2} f(p) dp = \frac{1}{2 m_{e}} \int_{p_{c}}^{\infty} p^{4} f dp \, ,
\label{enerdens}
\end{equation}
where $m_{e}$ is the electron mass, $\epsilon \equiv p^{2}/ 2 m_{e}$ is the electron kinetic energy. 
Consequently, the expression for the ratio of the nonthermal electron density ($n_{*}$) to the thermal one ($n_{e}$) is

\begin{eqnarray}
& {n_* \over n_e} =
{2 \, \xi_c^{(3+\alpha_1)/2} \, \Gamma\left(- {\alpha_1 \over 2} , \, \xi_c
\right) \over \sqrt{\pi} \, (3+\alpha_1) (\alpha_2-\alpha_1) \, D_0 \, \tau}
\nonumber \\
& \phantom{lotsofspaaace} + \, 2  \, e^{-\xi_c} \left(\xi_c \over \pi\right)^{1/2}
+ {\rm Erfc}\left(\xi_c^{1/2}\right)
\ ,
\label{eq6}
\end{eqnarray}
where $\xi \equiv \frac{p^{2}}{2m_{e}kT}$ and $\xi_{c}\equiv \frac{p_c^{2}}{2m_{e}kT}$.
The ratio of the energy density in the nonthermal electron population ($U_{*}$) to that in the thermal population ($n_{e} k T$) is

\begin{eqnarray}
& {U_* \over n_e k \, T} =
{2 \, \xi_c^{(5+\alpha_1)/2} \, \Gamma\left(- {\alpha_1 \over 2} , \, \xi_c
\right) \over \sqrt{\pi} \, (5+\alpha_1) (\alpha_2-\alpha_1) \, D_0 \, \tau}
\nonumber \\
& \phantom{lotsofspaaace} + \, {2 \sqrt{\pi \xi_c} \, (3 + 2 \, \xi_c) \, e^{-\xi_c}
+ \, 3 \, \pi \, {\rm Erfc}\left(\xi_c^{1/2}\right) \over
2 \, \pi (1 - 10 \, D_0 \, \tau)} \ .
\label{eq7}
\end{eqnarray}
Here T is the temperature of the thermal electrons, which we assume to be 1 MK. It is also instructive to calculate the brightness temperature $T_b$ of the noise storm radiation,
which is defined by
\begin{equation}
 T_b = 1.22 \times 10^{10} \frac{S}{f^2\theta\phi} \, ,
\end{equation}
where S is the flux in sfu, f is the observing frequency in GHz, and $\theta$ and $\phi$ are the source dimensions in arcseconds. This is essentially the ``equivalent'' blackbody 
temperature of the observed radiation in the Rayleigh-Jeans limit (which is applicable at radio frequencies). We take the highest
value in the light curve for $S$ and the smallest source dimensions from \citet{Mercier_2015}, so as to calculate the peak brightness temperature. The values are tabulated 
in \ref{tab:observed_para}. The high values of the brightness temperature (in comparison with the temperature of the solar corona, that is only as high as a few million K at most)
are indicative of the fact that the observed radiation is nonthermal.
 
\begin{figure}%
\subfloat{\includegraphics[scale=0.65]{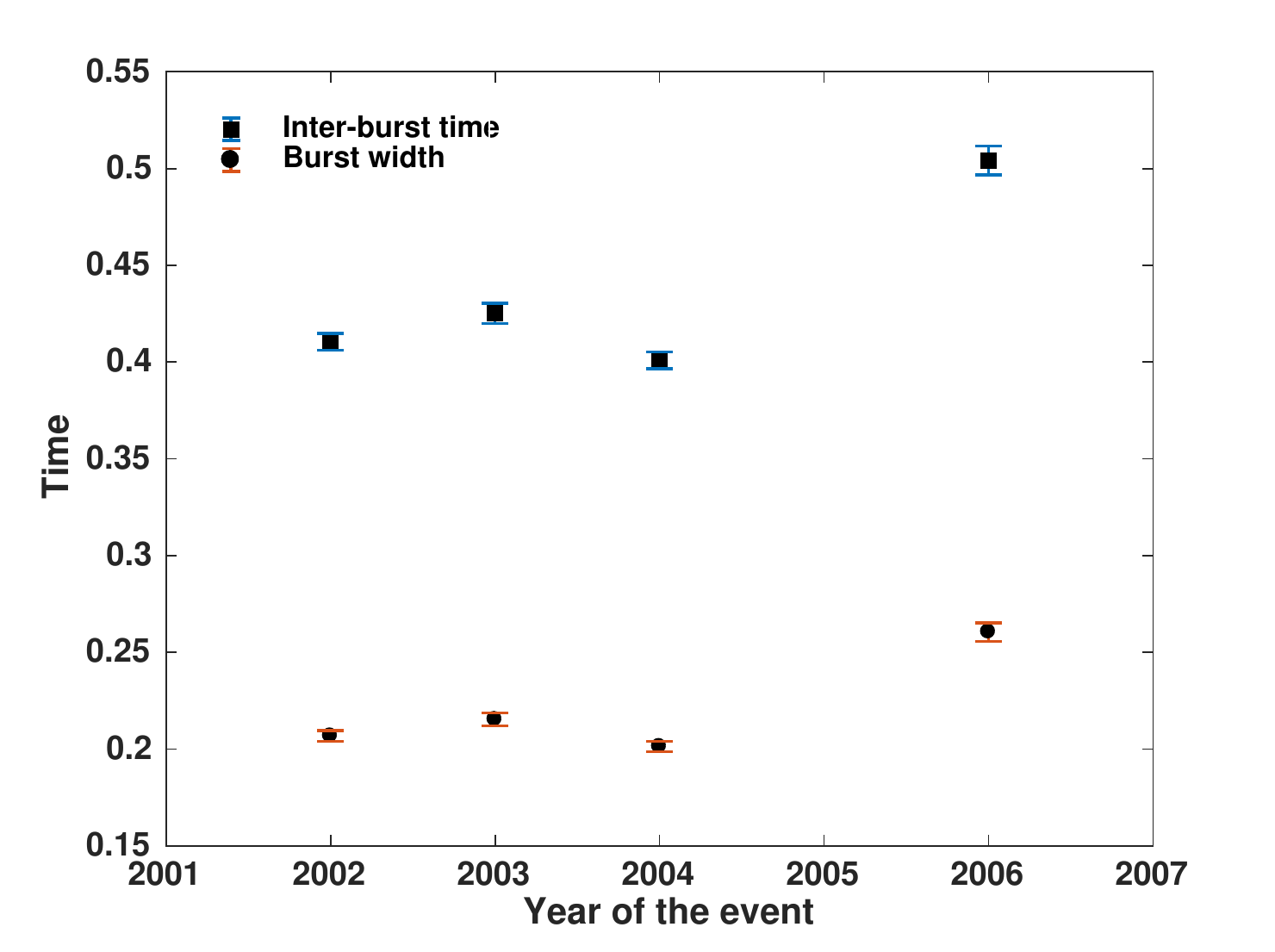}}%
  \caption{A scatterplot of burst width ($\tau$) and the inter-burst time interval ($t_{\rm accel}$). In all cases, $\tau/t_{\rm accel} < 4/5$}%
  \label{fig:burst_times}%
\end{figure}

\begin{table*}
	\centering
	\caption{Details of the noise storm events studied.} 
	
	\begin{tabular}{lcccc} 
		\hline
		Date    & Observation & Observation frequency            & Smallest source      & Derived Heliocentric \\
		        & timerange &  (MHz)          & size           &Distances ($R_\odot$) \\
		        &  (UT)      &                           & (arcsec)            & \\
		\hline
		2002 Aug.27 & 08:44-09:15& 327 & 31 & 1.20\\
		2003 Jul.15 & 10:44-11:15& 236 & 57 & 1.24\\
		2004 Aug.14 &11:39-12:15 & 327 & 45 & 1.21\\
		2006 Apr.06 &11:34-12:03 & 236 & 35 & 1.28\\
		\hline
	\end{tabular}
      \label{tab:event_list}
\end{table*}

\begin{table*}
	\centering
	\caption{Observed parameters of the noise storms.}
	\label{tab:parameters}
	\begin{tabular}{lccccc} 
		\hline
		Date & Number of  & Average            & Average      & Brightness &Powerlaw index \\
			&   bursts      &  duration of     &inter-burst   & Temperature & amplitude\\
			&  detected      & bursts (s)      & interval(s)  &  (K)  &   \\
		\hline
		2002 Aug.27 &4681 & 0.20 & 0.41  & 3.34x$10^{8}$&-3.86\\
		2003 Jul.15 & 3764 & 0.22 & 0.42 & 8.84x$10^{8}$&-3.76\\
		2004 Aug.14 & 4617  &0.21 & 0.40 & 6.37x$10^{8}$& -5.20\\
		2006 Apr.06 & 2968 & 0.27 & 0.50 & 1.24x$10^{10}$& -1.67\\
		\hline
	\end{tabular}
	\label{tab:observed_para}
\end{table*}

\begin{table*}
	\centering
	\caption{Derived parameters for the noise storm continua. }
	\label{tab:derived}
	\begin{tabular}{lcccccc} 
		\hline
		Date & $\alpha_{1}$  & $\alpha_{2} $  &$U_{*}/n_{e}kT$& $\frac{n_{*}}{n_{e}}$      & $U_{*}/n_{*}$ &$(2/3) \xi_c$\\
		& & & &   & & \\
                &               &        &        &                                                    &  (erg per electron)         & \\
		\hline
		2002 Aug.27 & 2.76 & -5.76 & 6.55x$10^{-7}$& 8.47x$10^{-9}$          &1.29x$10^{-8}$   & 13.26\\
		2003 Jul.15 & 2.66 & -5.66 & 2.76x$10^{-4}$& 5.32x$10^{-6}$          &8.71x$10^{-9}$   & 8.85\\
		2004 Aug.14 & 2.67 & -5.67 & 8.69x$10^{-7}$& 1.14x$10^{-8}$          &1.27x$10^{-8}$   & 13.05\\
		2006 Apr.06 & 2.65 & -5.65 & 6.91x$10^{-5}$& 1.10x$10^{-6}$           &1.01x$10^{-8}$   & 9.90\\
		\hline
	\end{tabular}
	\label{tab:derived_para}
\end{table*}

\subsection{Results and discussion}
\subsubsection{Noise storm continua}
For each noise storm episode, the mean value of the burst duration (column 3 of Table \ref{tab:observed_para} ) is identified with the mean residence time in the acceleration region ($\tau$).
The mean value of the inter-burst duration (column 4 of Table \ref{tab:observed_para}) is identified with the mean acceleration timescale ($t_{\rm accel}$). A scatterplot of $\tau$ versus
$t_{\rm accel}$ (Fig \ref{fig:burst_times}) shows that the ratio $\tau/t_{\rm accel} < 4/5$, as it should be, in order for the non thermal electron energy density $U_{*}$ to remain finite
\citep{Subramanian_2006}. Since the acceleration timescale is related to the quantity $D_{0}$ via Eq~\ref{eq5}, we can arrive at values for $\alpha_{1}$ and $\alpha_{2}$
(Eq~\ref{eq4}) for each event. These values are listed in columns 2 and 3 respectively of Table \ref{tab:derived_para}. Our knowledge of 
$\alpha_{1}$ and $\alpha_{2}$ enables us to calculate the ratio of the nonthermal electron density to the thermal one ($n_{*}/n_{e}$), the energy density of the nonthermal 
electron population ($U_{*}$), the ratio of the energy density in the nonthermal electron population to the thermal one ($U_{*}/(n_{e} k T)$) and the energy per electron
($U_{*}/n_{*}$) for each observed instance of the noise storm continuum. These quantities are listed in table \ref{tab:derived_para}. We recall that only electrons 
(from the thermal pool) beyond a critical energy are subjected to net acceleration. The quantity $(2/3) \xi_{c} \equiv (p_{c}^{2}/2 m_{e})/(3/2) k T_{e}$, also listed
in table \ref{tab:derived_para}, is a measure of how far removed this critical energy is from the energy characterizing the bulk thermal electrons. 
\subsubsection{Noise storm bursts}\label{sssec:num1}
We have discussed results pertaining to the noise storm continua so far. Assuming that the continuum is an aggregate of several individual noise storm bursts,
we now discuss properties of a representative individual burst.
We use the smallest observed source size for each event to 
calculate the volume ($V$) of the  noise storm emission region. In order to do this, we use $V = L^{3}$, where $L$ is the spatial extent corresponding to the source size in arcsec listed
in column 3 of Table \ref{tab:event_list}. This enables us to calculate the power supplied to the accelerated electron population ($P$). To find the energy ($E$) contained by the
electrons involved in a (representative) single burst using we multiply the power ($P$) with the total duration of the noise storm observation and divide by the number of bursts:
\begin{eqnarray}
\nonumber
P = V \frac{d U_{*}}{dt} =  t_{\rm accel}^{-1} \, V \, U_{*} \, \, \, \,\, {\rm erg\,s^{-1}} \, ,\\
E = P \, \frac{\rm duration \,of\,event}{\rm number \,of \,bursts} \,\,\,  {\rm erg}
\label{eq8}
\end{eqnarray}
The values of $P$ and $E$ for each noise storm episode are listed in column 2 and 3  of table \ref{tab:energy_rates}. It is instructive to compare the nonthermal energy density per burst with the corresponding energy density in the thermal population. The nonthermal energy density per burst is calculated by dividing the total energy per representative burst ($E$, column 3 of table \ref{tab:energy_rates}) by the volume ($V$) of the emission region. This quantity is listed in column 4 of table \ref{tab:energy_rates}. We assume that the ratio of nonthermal to thermal energy in the bursts is the same as that for the continuum (column 4 of table \ref{tab:derived_para}).  For each event, we divide the number in column 4 of table \ref{tab:energy_rates} by the corresponding number in column 4 of table \ref{tab:derived_para} to estimate the thermal energy density ``per burst''. Evidently, the thermal energy density exceeds the nonthermal energy density by three to seven orders of magnitude.

\begin{table*}
	\centering
	\caption{Parameters of a representative burst}
	\label{tab:derived}
	\begin{tabular}{lcccc} 
		\hline
		Date &  Power in electrons & Energy in accelerated                &Nonthermal energy                & Thermal energy\\
		     &  $({\rm erg \ s^{-1}})$     & electron population          & density per burst               &density per burst\\
		     &                                      &per burst(erg)            & (${\rm erg\,cm^{-3}}$ )    &(${\rm erg\,cm^{-3}}$ )\\

		\hline
		2002 Aug.27 &4.03x$10^{21}$ &7.24x$10^{20}$  &0.63x$10^{-7}$   & 0.10 \\
		2003 Jul.15 &3.35x$10^{24}$ &8.06x$10^{23}$  &1.69x$10^{-5}$   & 0.06\\
		2004 Aug.14 &1.67x$10^{22}$ &3.70x$10^{21}$  &1.06x$10^{-7}$   & 0.12\\
		2006 Apr.06 &2.61x$10^{23}$ &7.39x$10^{22}$  &4.52x$10^{-6}$    & 0.07\\
		\hline
	\end{tabular}
	\label{tab:energy_rates}
\end{table*}

\section{Conclusions}
Several numerical MHD studies suggest that the constant shuffling of field lines will lead to the formation of an ensemble
of small-scale current sheets and Alfvenic scatterers all over the solar corona \citep{Kumar2014,Kumar2016,Pisokas2018}; the number of current sheets per active region could be as
large as 100,000 \citep{Klimchuk_2015}. It is reasonable to expect that electrons interacting with magnetic scattering centers within such current sheets will lead to plasma heating and electron acceleration. This is the general expectation of coronal heating models that rely on the nanoflare scenario. However, 
there are few observationally based estimates of the energetics of such small electron acceleration episodes. We concentrate here on radio noise storms, which are 
the most commonplace instances of electron acceleration in the solar corona outside of large flares. They are typically observed above active regions, and last for several hours
to days. The noise storms we have observed comprise a train of spiky bursts with average durations of 0.20 to 0.27 seconds,if superposed on a background continuum.  The average inter-burst interval ranges from 0.4 to 0.5 seconds. We use high resolution imaging data for four noise storms observed jointly by the 
GMRT and the NRH, together with high temporal resolution data from the NRH. The high brightness temperatures of the emission ($\approx 10^{8}$ -- $10^{10}$ K) are suggestive of
an origin involving nonthermal/accelerated electrons. We use these data to constrain parameters of a model for stochastic acceleration of electrons 
\citep{Subramanian_2004,Subramanian_2006}. The inter-burst interval is interpreted as the electron acceleration 
timescale, while the burst duration is interpreted as the residence time of a typical electron in an acceleration region. The high resolution images provide information on the 
spatial extent of the noise storm emission region, which helps us arrive at an upper limit on the volume of the acceleration region.

We find that the occurrence frequency of the bursts is related to the burst amplitude via a power law, whose index ranges from -1.6 to -5.2 (table~\ref{tab:observed_para}). We find that the nonthermal electrons are a very small fraction of the thermal population in all cases - the fraction ranges from $10^{-9}$ to $10^{-6}$ (table~\ref{tab:derived_para}).
Similarly, the energy density in the nonthermal electron population  per burst is only $\approx$ $10^{-7}$--$10^{-3}$ times that in the thermal population.
This may be contrasted with the``cold'' small acceleration events studied by \citet{James_2017}, where the lack of soft Xray emission associated with their events implies
that the acceleration process produced mostly nonthermal electrons (and very few thermal ones).
We find that the power expended in accelerating the electrons responsible for a representative noise storm burst ranges from $10^{21}$ to $10^{24}$ ${\rm erg\,s^{-1}}$ and the 
energy contained in the accelerated electrons that produced a representative burst ranges from $10^{20}$ to $10^{23}$ erg (table~\ref{tab:energy_rates}). 
These numbers suggest that noise storm bursts (often called type I bursts) could be important contributors to active region coronal heating. It may be kept in mind that the numbers
quoted for the energy per nanoflare in most discussions refer to the amount of magnetic free energy released per nanoflare episode. This is distinct from the energetics of the 
nonthermal electron population responsible for such an episode, which is what we are concerned with in this work. Only some of the magnetic free energy is expended in accelerating and heating electrons, which eventually radiate. Since the efficiency of conversion from magnetic energy to nonthermal electrons is likely very small, we would expect the accelerated electron population to only contain a small fraction of the magnetic energy liberated in each nanoflare episode.

\section*{Acknowledgements}

We acknowledge a thorough review by Jim Klimchuk, which helped us improve the paper substantially. We acknowledge useful discussions with Divya Oberoi. TJ is thankful to IISER Pune for a PhD fellowship.




\bibliographystyle{mnras}
\bibliography{mnrasref} 








\bsp	
\label{lastpage}
\end{document}